\documentclass[preprint,showpacs,showkeys,preprintnumbers,amsmath,amssymb,aps]{revtex4-1}

\usepackage{bm,amssymb,amsmath,mathrsfs}
\usepackage[usenames]{color}

\usepackage{dcolumn}

\newcommand{\bc}{\begin{center}}
\newcommand{\ec}{\end{center}}
\newcommand{\bit}{\begin{itemize}}
\newcommand{\eit}{\end{itemize}}
\newcommand{\bq}{\begin{equation}}
\newcommand{\eq}{\end{equation}}

\begin{document}

\title{Comment on ``rf Wien filter in an electric dipole moment storage ring: The ``partially frozen spin'' effect''}

\author{S.~R.~Mane}
\email{srmane001@gmail.com}

\affiliation{Convergent Computing Inc., P.~O.~Box 561, Shoreham, NY 11786, USA}

\begin{abstract}
The Jacobi-Anger identity is employed to quantify the findings of
Morse, Orlov and Semertzidis \cite{Morse_Orlov_Semertzidis_2013}.
The secular term in the longitudinal spin component is proportional to a Bessel function.
A suggestion for values of machine parameters to optimize the secular term is presented.
\end{abstract}


\keywords{electric dipole moment, storage ring, Wien filter
{\em
\vskip 0.5in
\noindent
Morse, Orlov and Semertzidis published a paper in PRSTAB in Nov.~2013 \cite{Morse_Orlov_Semertzidis_2013}.
I submitted this comment to PRSTAB in Dec.~2013.
Following standard procedure, my manuscript was sent to the authors for an informal report (not an official Reply).
One of the authors (Morse) submitted an informal report
}
\vskip 0.15in
\noindent
REPORT OF THE INVOLVED REFEREE ZLK1002
\\
\noindent
The author is a very well-known physicist in this area. His Comment on
\\
\noindent
``rf Wien filter in an electric dipole moment storage ring: The
``partially frozen spin'' effect'' is interesting and relevant. I
recommend that it be published.
\vskip 0.15in
\noindent
{\em
However, the editorship of PRSTAB sent my manuscript for external review,
on the grounds that 
``\dots the information at hand is not an adequate basis for an editorial decision.''
An external referee is normally consulted only if no agreement is reached between the commenter
and the author(s) of the commented paper.
I terminated the submission of my manuscript in Dec.~2013.
This arXiv post is the utimate result.
}
}

\maketitle


\vskip 0.5in

\section{Introduction}
Morse, Orlov and Semertzidis \cite{Morse_Orlov_Semertzidis_2013}
have published an interesting idea to employ an rf Wien filter in an 
electric dipole moment (EDM) storage ring.
The basic idea is this (let us treat motion on the reference orbit only).
The particles circulate under the influence of static guide fields (a vertical magnetic and/or radial electric field).
The spin precession angular frequency under the influence of the guide fields
(in a frame corotating with the reference particle) is $\omega_g$, say.
In addition, the ring contains an rf Wien filter.
(In the simplest model, the Wien filter is uniformly distributed around the ring circumference.)
A Wien filter is a pair of crossed electric and magnetic fields such that the Lorentz force on the reference orbit vanishes:
$\bm{E}_{\rm wf} +\bm{\beta}_0\times\bm{B}_{\rm wf}=0$.
Wien filters have been employed as velocity filters and also as spin rotators in many accelerators,
see the review by Mane, Shatunov and Yokoya \cite{MSY2}.
Typically, Wien filters employ static electric and magnetic fields,
but an {\em rf Wien filter} is one where the electric and magnetic fields oscillate.
In particular, in the scenario described in \cite{Morse_Orlov_Semertzidis_2013},
the electric and magnetic fields oscillate {\em at the angular frequency $\omega_g$}, 
viz.~$\bm{E}_{\rm wf} = \bm{E}_{\rm wf0}\,\cos(\omega_g t)$ and
$\bm{B}_{\rm wf} = \bm{B}_{\rm wf0}\,\cos(\omega_g t)$,
where $E_{\rm wf} = \beta_0 B_{\rm wf}$.
This has the consequence that, if a particle is injected into ring with an initially longitudinal spin at $t=0$, 
the longitudinal spin component $s_L$ develops a {\em nonzero secular term}.
This is derived in \cite{Morse_Orlov_Semertzidis_2013}, and I shall explain it in more detail below.
It is pointed out in \cite{Morse_Orlov_Semertzidis_2013}
that this secular term in the longitudinal spin component can be used to furnish a testable EDM signal.
It is an interesting idea.

I shall now quantify the above ideas more mathematically.
A theoretical analysis (and results from tracking simulations) are of course presented in \cite{Morse_Orlov_Semertzidis_2013},
but the authors seem to be unaware of the Jacobi-Anger identity for Bessel functions,
which would simplify, and arguably provide greater insight into, some of their theretical derivations.
The secular term in the longitudinal spin component is (proportional to) a Bessel function.
A suggestion for values of machine parameters to optimize the secular term is presented.

\section{Secular term}
I treat a particle of mass $m$, charge $e$, velocity $\bm{\beta}c$ and spin $\bm{s}$ moving in external fields $\bm{E}$ and $\bm{B}$.
The Lorentz factor is $\gamma=1/\sqrt{1-\beta^2}$ and the magnetic moment anomaly is $a=(g-2)/2$.
I shall follow \cite{Morse_Orlov_Semertzidis_2013} and treat a particle on the reference orbit only.
The independent variable is the time $t$.
A positive bend is to the right (i.e.~clockwise).
The radial, longitudinal, and vertical components of the spin are denoted by $s_R$, $s_L$ and $s_V$ respectively.
The spin precession equation on the reference orbit (where $\bm{\beta}\cdot\bm{B} = \bm{\beta}\cdot\bm{E} = 0$) 
is, from \cite{Morse_Orlov_Semertzidis_2013}
\begin{align}
\frac{d\bm{s}}{dt} &= \omega \times \bm{s} \,,
\\
\bm{\omega} &= \bm{\omega}_a + \bm{\omega}_{\rm edm} \,,
\\
\bm{\omega}_a &= -\frac{e}{mc}\,\biggl[\,a\bm{B} + \biggl(a-\frac{m^2c^2}{p^2}\biggl)\bm{E}\times\bm{\beta}\,\biggr] \,,
\\
\bm{\omega}_{\rm edm} &= -\frac{e\eta}{2mc}\,(\bm{E}+\bm{\beta}\times\bm{B})\,.
\end{align}
Here $\eta$ is a measure of the EDM, viz.~for the electric and magnetic dipole moments
\bq
d = \frac{\eta}{2}\,\frac{e}{mc}\,s \,,\qquad
\mu = \frac{g}{2}\,\frac{e}{mc}\,s \,.
\eq
The authors in \cite{Morse_Orlov_Semertzidis_2013} 
do not state that the above expression for $\bm{\omega}$
is in a frame comoving with the reference particle, and {\em not} in a fixed reference frame.

The `unperturbed' model consists of a vertical magnetic field $B_V$ and a radial electric field $E_R$.
In \cite{Morse_Orlov_Semertzidis_2013} the authors seem to treat an all-magnetic guide field (i.e.~$E_R=0$) 
for most of their derivation (until they treat the EDM motion in Secion III in \cite{Morse_Orlov_Semertzidis_2013})
but there is no need to do so.
I shall allow $E_R\ne0$ throughout.
To avoid confusion of notation I define the spin precession vector due to the guide field as
\bq
\bm{\omega}_g = -\frac{e}{mc}\,\biggl[\,aB_V - \biggl(a-\frac{m^2c^2}{p^2}\biggl)\beta_0E_R\,\biggr]\,\bm{e}_V \,.
\eq
Here $\bm{e}_V$ is a unit vector in the vertical direction (and $\bm{e}_R$ and $\bm{e}_L$ are defined with an obvious analogy).
The spin precession vector due to the rf Wien filter is
\bq
\begin{split}
\bm{\omega}_{\rm wf} &= -\frac{e}{mc}\,\biggl[\,aB_{\rm wf0} - \biggl(a-\frac{1}{\gamma_0^2-1}\biggl)\beta_0E_{\rm wf0}\,\biggr]\,
\cos(\omega_gt) \,\bm{e}_V 
\\
&= -\frac{e}{mc}\,\frac{1+a}{\gamma_0^2}\,B_{\rm wf0} \,\cos(\omega_gt) \,\bm{e}_V 
\end{split}
\eq
We neglect $\bm{\omega}_{\rm edm}$ as a small perturbation.
Then the spin precession vector 
$\bm{\omega} = \bm{\omega}_a+\bm{\omega}_{\rm edm} 
\simeq \bm{\omega}_a = \bm{\omega}_g+\bm{\omega}_{\rm wf}$ 
is vertical and the spin precesses in the horizontal plane.
Then
\bq
\frac{d\ }{dt}\begin{pmatrix} s_L \\ s_R \end{pmatrix} =
\begin{pmatrix} 0 & -\omega_a \\ \omega_a & 0 \end{pmatrix} 
\begin{pmatrix} s_L \\ s_R \end{pmatrix} \,.
\eq
The solution at time $t$, starting from $t=0$, is
\bq
\begin{pmatrix} s_L \\ s_R \end{pmatrix} =
\begin{pmatrix} \cos\Phi & -\sin\Phi \\ \sin\Phi & \cos\Phi \end{pmatrix} 
\begin{pmatrix} s_L \\ s_R \end{pmatrix}_{t=0} \,.
\eq
Here the spin rotation angle is 
\bq
\Phi = \int_0^t \omega_a(u) \,du = \omega_g t -\frac{eB_{\rm wf0}}{mc\,\omega_g}\,\frac{1+a}{\gamma_0^2}\, \,\sin(\omega_gt) 
\equiv \omega_g t -\xi \,\sin(\omega_gt) \,.
\eq
This is where the Bessel funtions come in. 
The Jacobi-Anger identity for Bessel functions is
\bq
e^{ir\sin\psi} = \sum_{n=-\infty}^\infty e^{in\psi}J_n(r) \,.
\eq
Hence
\bq
e^{i\Phi} = e^{i\omega_gt} \, \sum_{n=-\infty}^\infty e^{-in\omega_gt}J_n(\xi) \,.
\eq
Then a little algebra yields
\begin{align}
\cos\Phi &= J_1(\xi) + \sum_{n=1}^\infty \bigl[\,J_{1-n}(\xi)+J_{1+n}(\xi)\,\bigr]\,\cos(n\omega_gt) \,,
\\
\sin\Phi &= \phantom{J_1(\xi) +} \sum_{n=1}^\infty \bigl[\,J_{1-n}(\xi)-J_{1+n}(\xi)\,\bigr]\,\sin(n\omega_gt) \,.
\end{align}
Both $\cos\Phi$ and $\sin\Phi$ contain an infinite sum of Fourier harmonics at integer multiples of $\omega_g$.
In addition, the value of $\cos\Phi$ has a nonzero secular term.
For initial conditions $s_L=1$ and $s_R=0$ at $t=0$, 
we obtain $s_L(t) = \cos\Phi$ and $s_R(t) = \sin\Phi$.
Then the longitudinal spin component $s_L(t)$ has a nonzero secular term, hence its time average is nonzero.
The time average of $s_L(t)$ is
\bq
\overline{s_l(t)} = J_1\Bigl(\frac{eB_{\rm wf0}}{mc\,\omega_g}\,\frac{1+a}{\gamma_0^2}\Bigr)
\simeq \frac{eB_{\rm wf0}}{2mc\,\omega_g}\,\frac{1+a}{\gamma_0^2} \,.
\eq
The expression derived in \cite{Morse_Orlov_Semertzidis_2013} is the leading order expansion of $J_1(\xi)$,
as displayed in the approxmation above.

\section{EDM signal}
Next, it is pointed out in \cite{Morse_Orlov_Semertzidis_2013} how the above result can be employed 
to derive a testable signal for a nonzero EDM.
For this we analyze the evolution of the vertical spin component.
Since $\bm{\omega}_g$ is vertical and $\bm{\omega}_{\rm edm}$ is radial,
only $\bm{\omega}_{\rm edm}$ contributes to the time evolution of $s_V$, viz.
\bq
\frac{ds_V}{dt} = -\omega_{\rm edm} s_L 
= \frac{e\eta}{2mc}\,(E_R-\beta_0B_V) \,s_L(t) \,.
\eq
This has the opposite sign to the expression in \cite{Morse_Orlov_Semertzidis_2013}, 
which may be due to a difference of coordinate systems.
As noted in \cite{Morse_Orlov_Semertzidis_2013}, 
only the guide fields appear in $\bm{\omega}_{\rm edm}$ 
because by definition the Wien filter terms in $\bm{\omega}_{\rm edm}$ 
cancel to zero: $\bm{E}_{\rm wf} +\bm{\beta}_0\times\bm{B}_{\rm wf}=0$.

We see that the time rate of change of $s_V$ has a nonzero secular term.
Hence a nonzero EDM will generate a secular rate of rotation of the spin out of the horizontal plane.
Taking a time average, we obtain 
\bq
\begin{split}
\overline{\frac{ds_V}{dt}} &= \frac{e\eta}{2mc}\,(E_R-\beta_0B_V) \,\overline{s_L(t)}
\\
&\simeq \eta\,\frac{eB_{\rm wf0}}{mc}\,\frac{1+a}{\gamma_0^2} \,\frac{e(E_R-\beta_0B_V)}{mc\, \omega_g} \,.
\end{split}
\eq
This is the expression derived in \cite{Morse_Orlov_Semertzidis_2013}, up to a global minus sign.
The expression in \cite{Morse_Orlov_Semertzidis_2013} contains a normalization factor $s_{L0}$
but I have normalized the spin to a unit vector.
The authors in \cite{Morse_Orlov_Semertzidis_2013} go on to analyze various systematic errors to the EDM signal,
which I do not treat here.

\section{Summary: the `partially frozen spin' method}
One technique to detect an EDM signal is the so-called `frozen spin' method.
This requires the longitudinal spin component to be constant in time.
(This is attained by setting $\omega_a=0$, so that $s_L$ does not vary in time.)
Then $ds_V/dt$ exhibits a nonzero time average, as indicated above.
The technique described in \cite{Morse_Orlov_Semertzidis_2013} is a so-called `partially frozen spin' method
in the sense that the longitudinal spin component is not completely constant in time, 
but it {\em does} contain a nonzero secular term.
It might be best (if feasible) to configure the machine parameters so that $J_0(\xi)\simeq 0$, i.e.~$\xi\simeq 2.4$,
so as to minimize the unwanted nonsecular terms
(in which case the secular term $J_1(\xi)$ is also close to its maximum value).
It is an interesting idea.


\end{document}